# Visualizing topological edge states of single and double bilayer Bi supported on multibilayer Bi(111) films

Lang Peng[1], Jing-Jing Xian[1], Peizhe Tang[2,3,†], Angel Rubio[3], Shou-Cheng Zhang[2,4], Wenhao Zhang[1,*], Ying-Shuang Fu[1,#]

1. School of Physics and Wuhan National High Magnetic Field Center, Huazhong University of Science and Technology, Wuhan 430074, China
2. Department of Physics, McCullough Building, Stanford University, Stanford, California 94305-4045, USA
3. Max Planck Institute for the Structure and Dynamics of Matter, Luruper Chaussee 149, 22761, Hamburg, Germany
4. Stanford Institute for Materials and Energy Sciences, SLAC National Accelerator Laboratory, Menlo Park, California 94025, USA
   Emails: *wenhaozhang@hust.edu.cn, #yfu@hust.edu.cn, †peizhe.tang@mpsd.mpg.de

**Freestanding single-bilayer Bi(111) is a two-dimensional topological insulator with edge states propagating along its perimeter. Given the interlayer coupling experimentally, the topological nature of Bi(111) thin films and the impact of the supporting substrate on the topmost Bi bilayer are still under debate. Here, combined with scanning tunneling microscopy and first-principles calculations, we systematically study the electronic properties of Bi(111) thin films grown on a $NbSe_2$ substrate. Two types of non-magnetic edge structures, i.e., a conventional zigzag edge and a 2×1 reconstructed edge, coexist alternately at the boundaries of single bilayer islands, the topological edge states of which exhibit remarkably different energy and spatial distributions. Prominent edge states are persistently visualized at the edges of both single and double bilayer Bi islands, regardless of the underlying thickness of Bi(111) thin films. We provide an explanation for the topological origin of the observed edge states that is verified with first-principles calculations. Our paper clarifies the long-standing controversy regarding the topology of Bi(111) thin films and reveals the tunability of topological edge states via edge modifications.**

## I. INTRODUCTION

Topological insulators (TIs) represent a class of materials that possess an insulating interior and metallic surface/edge states at the boundary. These surface/edge states are topologically protected by time-reversal symmetry and their gapless properties are robust against non-magnetic perturbations, which can induce applications in energy-efficient electronic devices [1,2]. For two-dimensional (2D) TIs, the topological edge states (TESs) on the one-dimensional (1D) boundary will contribute to quantum spin Hall effect with a longitudinal conductance of $2e^2/h$ [3,4]. A pioneering system hosting the 2D TI phase is graphene, where hopping between its two sublattices creates Dirac fermions, and the spin-orbit coupling (SOC) adds a massive term to its Dirac Hamiltonian and concomitantly drives it into the topological regime [5]. However, its SOC strength is too weak to allow experimental examination [6,7]. Bi(111) thin film has a layered structure. Each layer has a buckled honeycomb lattice, forming a bilayer (BL) structure. Due to its structural similarity with graphene and huge SOC, Bi(111) thin film has sparked extensive research interests in studying its topological phase.

A single-BL Bi(111) is theoretically predicted to be a 2D TI [8-10] and experimentally confirmed by probing the TESs in such films supported on three-dimensional TI substrates [11-14], as well as by measuring quantum transport in mechanically exfoliated films [15]. On the other hand, the situation becomes complicated for the multi-BLs. If the inter-BL coupling of Bi(111) thin film is negligible, its topology will exhibit an odd-even oscillation as a function of film thickness $N$ with odd $N$-BL film being topological, as has been predicted in the early studies [8], whereas subsequent theoretical investigations demonstrate that the inter-BL coupling should be considered in reality. As a result, rather than the odd-even oscillation, the Bi(111) film is globally topological up to 8 BL [16], above which a topological phase transition occurs [17]. Experimentally, spin-resolved angle-resolved photoemission spectroscopy (ARPES) studies on a 15 BL Bi film detect the 1D edge states with spin-split bands of large Rashba type [18], implying that a Bi thin film with edge structures is topologically trivial as the bulk Bi [19,20]. Subsequent ARPES data, however, prove the presence of topologically protected surface states in 14-202 BL Bi films, suggesting their nontrivial topology [21,22]. Recent scanning tunneling microscopy (STM) studies also report the observation of nontrivial TESs at single-BL edges supported on bulk Bi and 96 BL Bi films [23,24]. These controversial results on the topological nature of Bi films suggest that the coupling with Bi substrate could be a factor to influence the topological nature of single-BL Bi [10,18]. Therefore, it is essential to scrutinize the edge states of single-BL Bi on various thicknesses of Bi thin films underneath, especially the relationship between the TESs and the underlying substrates.

In addition, the edge structure of Bi films also plays a vital role in tuning its electronic structures. On the bare zigzag edge of freestanding single-BL, the unsaturated dangling bonds of Bi atoms tend to form anti-ferromagnetism through direct coupling that will break the time-reversal symmetry. These magnetic zigzag edges are analogy to those in graphene [25]. With the presence of substrates, dangling bonds will interact with the substrate strongly, which could destroy the magnetism on edges and modify their electronic properties [26,27]. Previous STM measurements indicate that only a portion of the topological signature can be observed on the boundaries of single-BL Bi supported on Bi crystal, where the TES is suppressed when the edge atoms strongly couple with the Bi substrates [23]. More interestingly, theoretical calculations also predict that the characteristic features of TESs on a single-BL Bi(111) edge, such as the Fermi velocity and spin textures, will be significantly tuned via the adsorption of hydrogen atoms [28-30]. To this end, it is indispensable to systematically investigate the edge structures of Bi(111) by considering coupling with substrates, as well as their consequent impact over the TESs.

In this paper, we report the spectroscopic visualization of 1D edge states of both single- and double-BL Bi(111) islands supported on Bi thin films. Utilizing STM and scanning tunneling spectroscopy (STS) measurements and density functional theory (DFT) calculations, we uncover two types of non-magnetic edges on the single-BL Bi islands: a conventional zigzag edge and a reconstructed edge with 2×1 periodicity. Both types of edges reveal topologically nontrivial edge states, which distribute in a very different energy range and spatial extension. By changing the thickness of the underlying Bi substrate (from 4 to 9 BL), we can always observe these states although their electronic properties vary a lot. To understand the topological origin of the observed edge states, a unified scenario is built by using the perturbation theory. Our findings unambiguously verify the robustness and tunability of TESs in single-BL Bi(111) islands against the supporting film thickness or structural edge decorations.

## II. METHODS

The $NbSe_2$ single crystal is cleaved in ultrahigh vacuum at 90 K as substrates with a base pressure of $2\times10^{-10}$ Torr. High purity of Bi (99.999%) is evaporated from a standard Knudsen cell at 723 K. The $NbSe_2$ substrate is held at room temperature and the growth rate is ~ 1.5 Å/min. After growth, the Bi film is post-annealed at 520 K for 30 min to obtain flat surfaces. The experiments are conducted with a Unisoku STM system at 4.5 K. Normal W tips are cleaned by electron-beam heating and checked on Ag islands before the STM and STS measurement. All topographic images are taken in constant-current mode, and the tunneling d$I$/d$V$ is recorded by standard lock-in technique with modulation voltage frequency of 983 Hz and amplitude of 10 mV if not specifically noted.

The *ab initio* calculations are carried out in the framework of DFT with the projector augmented wave method [31], as implemented in the Vienna *ab initio* simulation package [32]. A plane-wave basis set is used with a kinetic energy cutoff of 200 eV and the generalized gradient approximation functional is used as the exchange-correlation functional [33]. The in-plane lattice constant is used from experimental values, and is fixed as 0.454 nm. The freestanding single-BL nanoribbons contain 14 or 15 zigzag chains [27] and the in-plane vacuum layer is 10 Å, such that the distance between neighboring edges is large enough to avoid interaction between edge states. Along the *z* direction, the vacuum layer is 15 Å. Atomic positions in the nanoribbons are allowed to fully relax until residual forces are less than $1\times10^{-3}$ eV/Å. For single-BL nanoribbons on Bi(111) thin films, we put the relaxed nanoribbons on the Bi(111) substrate and make sure that the distance between nanoribbons is large enough to avoid their coupling. Then we relax the entire system to find the stable structure. In this model, the substrate with the single-BL Bi(111) is used and the coupling between single-BL nanoribbon and the substrate is fully considered, but the quantum confinement of the substrate may be overestimated. The Tersoff-Haman approximation is used to simulate the STM images [34,35], in which the local density of states is integrated from -0.5 eV to the Fermi level ($E_F$). Because of the reconstruction of the edges, the edge Bi atoms on the outermost layer are not in the same plane with those in the bulk. Herein, we focus on the edge, the distance between the simulated STM image and the centers of mass for edge Bi atoms on the outermost layer is about 2 Å. In these calculations, the Monkhorst-Pack k points are 15×1×1, and SOC is included in the self-consistent calculations. For the nanoribbon, the formation energy is defined as

$E_F = (E_{total} - E_{atom} \times N)/2L$, in which $E_{total}$ is the calculated total energy for the nanoribbons, $E_{atom}$ is the energy per atom for single-BL Bi(111), $N$ is the number of Bi atoms in the nanoribbons, and $L$ is the length of the edge in one unit cell. In these formation energy calculations, the SOC effect is not included.

## III. RESULTS AND DISCUSSION

### A. Surface topography and structural geometry

Previous theoretical and experimental studies indicate that there are two distinctive allotropic structures for epitaxial Bi ultrathin films: one is Bi(111), a hexagonal facet in the (111) orientation of rhombohedral bulk Bi. The other is Bi(110), a pseudocubic facet in the (110) orientation. Figure 1(a) presents a typical STM topographic image of Bi film grown on $NbSe_2$ substrate. With the help of atomic resolution images (see inset), we experimentally find that our epitaxial Bi film adopts (111) orientation only when the thickness is larger than 4 BL, below which Bi forms (110)-oriented thin film. This structural transformation is consistent with Bi films grown on other substrates previously [36,37]. The STS spectra (Fig. 1(c)) taken on different surfaces in Fig. 1(a) suggest that the electronic structure of Bi thin film is strongly thickness dependent, different from bulk Bi form [23].

The Bi(111) film is composed of buckled honeycomb BLs, the sublattice atoms of which in a BL are illustrated in Fig. 2(a) with cyan and magenta colors, respectively. Owing to the different interaction between the sublattice and substrate, there are two types of edge terminations discovered in our experiments. The different edges can be seen in our STM image of a Bi(111) thin film with a single-BL-height pit (Fig. 2(b)), the shape of which is a hexagon constituted of alternating longer and shorter edges. The different edge character is carefully examined by our high-resolution STM images in Figs. 2(c) and 2(d). While the lattice spacing of the longer edge is the same as that of the bulk (0.45 nm, inset of Fig. 2(b)), consistent with previous reports [23,24], the shorter edge displays a 2×1 reconstruction with the adjacent atom spacing of 0.52 and 0.38 nm (Fig. 2(e)), respectively. A similar 2×1 reconstruction is observed in Bi films prepared on the Si(111) substrate before [24], but is missing in cleaved bulk samples [23]. Hereafter, we name the conventional zigzag-type edge and the reconstructed edge as the Z-edge and R-edge, respectively.

To understand the origin of the observed new edge structure (R-edge), we calculate the structural and electronic properties for single-BL Bi(111) nanoribbons by using the DFT calculations. First, we fully optimize the lattice structures of free-standing zigzag nanoribbons with different edge geometries and calculate their edge formation energies. As shown in Fig. 3(a), we find a stable non-magnetic edge structure with 2×1 reconstruction (R-edge). Due to the dimerization, the dangling bonds on the edge are fully saturated, thus the edge formation energy of the R-edge (0.523 eV/Å) is much lower than that of the bare zigzag edge without relaxation (0.729 eV/Å, following the terminology previously used in the field of graphene [38] as the perfect Z-edge). Figure 2(f) shows the calculated band structure for freestanding nanoribbons with R-edges. Shadowed regions mark its bulk bands and red lines show the nontrivial edge states. The TESs couple with the trivial flat band, and the Dirac-cone characteristic can be clearly observed around $E_F$. Then, we put the ribbon with edge reconstruction on a BL Bi(111) substrate, the most stable structure of which is shown in Fig. 2(h). Although its dimerized edge atoms couple with the underlying substrate, its R-edge is still non-magnetic.

Then, we explore the structural and electronic properties for the Z-edge. Figure 3(b) shows the geometric structure of the freestanding ribbons with perfect Z-edge. Due to the direct coupling between unpaired dangling bonds, the antiferromagnetic order can be always observed in such a system whether the SOC effect is considered or not. Then we fully relax the ribbon with perfect Z-edge to achieve a more stable structure shown in Fig. 3(c), the edge formation energy of which is 0.609 eV/Å. Different from the case with the R-edge, we do not observe the dimerization here but find an antiferromagnetic order when the SOC effect is not considered (Fig. 3(e)). When SOC is considered in the calculation, we cannot observe the magnetism, which indicates that the strong SOC effect in Bi(111) can quench magnetic moments on the edges. We plot band structures for freestanding ribbons with relaxed Z-edge in Fig. 3(d). For the case without SOC, we observe the in-gap states that are marked by the dotted lines in Figs. 3(e) and 3(f). Via plotting the charge distribution for these in-gap states in Figs. 3(g) and 3(h), we find that they are mainly localized on the edge atoms. When SOC is considered, the topological nontrivial edge states can be observed around $E_F$. Finally, we put the ribbon with relaxed Z-edge on the substrate, relax the whole structure, and calculate its electronic structure with SOC. The most stable structure is shown in Fig. 2(g), in which the magnetism is absent due to the strong coupling between dangling bonds and Bi(111) substrates.

The calculated results of both types of edges being non-magnetic are substantiated by our STM measurements. Considering our Bi films are grown on a superconducting (SC) NbSe$_2$ substrate, they are subject to proximity effect, and prominent SC gaps are observed on all Bi films. In Fig. 4, we find the SC gap is undisturbed at both Z- and R-edges and nearby, with no visible in-gap states observed along and across the edges. Since the conventional s-wave superconductivity in NbSe$_2$ is very sensitive to local magnetic order, the above observation demonstrates the absence of magnetism on both Z- and R-edges [39,40]. Further simulated STM images (Figs. 2(i) and 2(j)) for both types of edges are qualitatively consistent with our experimental observations. We note that the underlying Bi atoms near the edge deviate from the perfect honeycomb positions (gray thin lines in Figs. 2(g) and 2(h)) when the top edge Bi atoms are arranged in a zigzag or reconstructed geometry, different from the type-A and type-B edges in Ref. [23].

### B. Topological edge states propagating along the Bi(111) perimeter

To unravel the response of the edge states to the different atomic edge structures, we perform detailed spatial dependent STS measurements around the step edges of a single-BL Bi island on the 7 BL substrate (Fig. 5(a)). As shown in Fig. 5(b), the spectrum exhibits a strong peak at 238 and 233 mV, respectively, on the terrace of 7 BL (magenta curve) and 8 BL (red curve). This peak, ascribed to van Hove singularities (VHSs) of the Bi band structure [41,42], locates at a slightly larger energy than that of the bulk value (213 mV) reported previously [23,43]. The VHS peak (upward arrows in Fig. 5(b)) is spatially uniform over the whole terrace but exhibits energy shift near the single-BL edges. For the Z-edge, the VHS peak moves to 198 mV for both upper and lower edges of the step. As is more clearly seen in the 2D plot of tunneling spectra measured perpendicularly across the edge (Fig. 5(c)), besides the quasiparticle interference (QPI) patterns originated from the 2D surface states of Bi(111) thin film, a remarkably localized state is found exactly at the upper edge within the confines of 3.5 nm (Fig. 6(a)). Such a state is homogeneously distributed along the Z-edge (Fig. 5(e)) and does not move away from the edge as a function of energy from 150 to 220 mV, suggesting an obvious signature of 1D edge state.

In contrast, the edge state is quite different for the R-edge (dashed curves in Fig. 5(b)). Therein, the VHS peak shifts to 184 mV for the lower edge but is strongly suppressed for the upper edge. Meanwhile, a broad but explicit peak emerges near 100 mV at the upper edge (downward arrow).

Despite the disparate features from the Z-edge, the 1D edge state still presents in the range of energies from $E_F$ to 120 mV, also right at the upper R-edge (Fig. 5(d)), but with a narrower spatial extension of 1.5 nm in the real-space distribution (Fig. 6(b)). Interestingly, the edge state along the R-edge exhibits a conductance modulation of 0.9 nm (Fig. 5(f)), which is in accordance with the 2×1 edge reconstruction. We note that the 0.9-nm reconstruction on the R-edge is independent of scanning bias in a wide energy range, which is sharply different from the Z-edge (Figs. 7(a)-(d)).

We distinguish the edge states on the Z- and R-edge from QPI standing waves in three aspects. First, the edge states are energy dispersionless while the standing waves are energy dependent. Second, the 1D states are exactly at the upper step edge (Fig. 5(b)), but the location of standing waves varies at different energies in real space. Third, we perform the 1D Fourier transformation of the STS line cut along the Z-edge to show the energy dispersion of the 1D states. Despite the poor resolution in Fig. 6(c), we can still resolve two scattering channels with high intensities (marked with $q_1$ and $q_2$) that disperse to lower energies with increasing the wave number (similar $q_1'$ and $q_2'$ in Fig. 6(d)). These features exhibit some similarities in scattering properties with previous studies [23,24], where the scattering wave vector $q_2$ originates from the 2D surface states, while $q_1$ has been identified as edge states of topological origin. The suppressed signals of $q_1$ at lower energies reflect that scattering only occurs between the states of similar spin, but backscattering is strongly suppressed.

On the other hand, our calculated band structures for the freestanding single-BL Bi(111) nanoribbon (Figs. 2(f) and 3(d)) indicate that the TESs develop within the energy range of the 2D bulk energy gap, reflecting the bulk properties of single-BL islands. However, different from the freestanding case, the bulk states of single-BL islands are no longer fully gapped due to strong interlayer coupling with the supporting Bi films, resulting in the non-insulating behaviors in our STS observations. The edge states located within the energy window of the bulk band gap, exhibiting nonzero STS intensities, have been also reported previously [12,23]. This in-gap property of edge state is consistent with the suppression of the backscattering wave vector in the 1D Fourier transform of STS (Figs. 6(c) and 6(d)), as discussed in Ref. [23]. Overall, our experimental characterizations of edge states at the Z- and R-edge, including the 1D localized states (Figs. 5(c) and 5(d)), the weakly visible scattering dispersions (Figs. 6(c) and 6(d)), and the suppressed STS intensities of bulk states near the step edge (Figs. 5(b)), indicate these edge states to be topological, qualitatively in accordance with our theoretical calculations.

As predicted theoretically, the well-localized 1D TESs persisting along the perimeter of single-BL Bi boundaries, however, may be significantly modified in detailed electronic structure [28-30]. This expectation is validated by our spectroscopic visualization of Z- and R-edge states with different energy range and spatial extension (Figs. 5 and 6), indicating the feasibility of tuning the TESs by edge modification. In contrast to the case of single-BL on bulk Bi [23,24], the reconstruction at the R-edge partially releases the coupling with the underlying substrate, which makes the TESs detectable and allows the TESs to propagate along the edge of a single-BL. As is exemplified in a single-BL pit containing both Z- and R-edges (Fig. 7(e)), the alternating appearance of edge states for the R-edge at 100 mV and Z-edge at 200 mV confirms the existence of an edge mode propagating along the hexagonal perimeter (Figs. 7(f) and 7(g)).

### C. Topological scenario for various thicknesses of Bi films

Next, we investigate the evolution of the single-BL TESs as a function of the underlying Bi(111) film thickness, starting from 4 BL (Fig. 1). As shown in Fig. 8(a), by increasing the thickness of Bi films up to 10 BL, the VHS peak on the Bi(111) terrace monotonously shifts towards lower energy, and finally becomes the same value as bulk Bi (213 mV, dashed black line in Fig. 8(c)). Accordingly, the state at the Z-edge of the single-BL step (Fig. 8(b)) also evolves with the same thickness dependence, as depicted by the red curve in Fig. 8(c). Nevertheless, we can always observe the 1D edge states localized at the Z-edge of the single-BL island. It should be noted that the critical thickness of topological phase transition (8 BL) for the whole thin film has been covered in our observations [16,17]. For instance, Figs. 8(d) and 8(e) show similar signatures of edge states with Figs. 5(c) and 5(d) supported on a 9 BL substrate, which are marked by the dashed magenta and black lines for the Z- and R-edge, respectively. Here, the downwards shifted VHSs (Figs. 8(a) and 8(b)) at both Z- and R-edges result in localized states distributed within the STS region of suppressed intensities, i.e. inside the bulk band gap, although the energy range is different from Figs. 5(c) and 5(d). Per the aforementioned discussion, we thus conjecture that the TESs of single-BL still persist for a thicker Bi substrate (9 BL). These results contradict previous theoretical prediction that single-BL Bi(111) on a sufficiently thick film will evolve into a trivial 2D metal [17], but imply that the topological nature of the single-BL Bi(111) is barely affected by the supporting Bi film substrate.

To justify the experimental implications, we apply the perturbation theory to study the topological properties of top single-BL Bi and the entire system by varying the thickness of Bi substrates. First of all, we would like to clarify one point: the existence of the TESs on the single-BL only reflects the nontrivial bulk topology of the top single-BL Bi, although its electronic properties can be influenced by the interlayer coupling with the underlying Bi(111) substrate [8,16]. In this sense, when discussing the topological origin of the TESs, we need to focus on the top single-BL rather than the entire system. This scenario is different from previous discussions [17].

With this notion, we divide the system into two parts as shown in Fig. 9(a): the top single-BL Bi(111) where the TES is observed, and the underlying multi-BL Bi(111) film substrate. The band evolutions for the states around $E_F$ are schematically shown in Figs. 9(b) and 9(c). The cyan (red) lines stand for the conduction- and valence-band edge at the $\Gamma$ point for top single-BL (film substrate) in isolation, marked as "CBE@$\Gamma$" and "VBE@$\Gamma$". Before we consider the interlayer coupling, both top single-BL and multi-BL substrate are 2D TIs. The topologically nontrivial band gap in the top single-BL is much larger than that in the multi-BL substrate due to quantum confinement. The interlayer coupling, which is represented by orange and green dashed lines in Figs. 9(b) and 9(c), could possibly to induce a topological phase transition in the whole system, depending on the thickness of the Bi substrate.

As calculated in Fig. 9(d), all Bi(111) thin films with different thicknesses always have an energy gap at the $\Gamma$ point around $E_F$. For the system containing the multi-BL substrate <7 BL ( the total film thickness should be <8 BL including the top single-BL), the interlayer coupling can induce a band inversion around $E_F$, keeping the entire system in the topological phase (Fig. 9(b)). However, for that with a thicker substrate (>8 BL), the entire system is a trivial semimetal (Fig. 9 (c)). This conclusion is consistent with previous work [16].

On the other hand, if we just focus on the electronic structure of the top single-BL instead of the entire system when the interlayer coupling is considered, its topological property does not change when increasing the thickness of underlying substrates (see cyan lines in the middle panel in Figs. 9(b) and 9(c) for band edges). Even when the global system is topologically trivial, the top single-BL still has the same band order with freestanding single-BL (Fig. 9(c)). That is to say, the edge states in top single-BL are always detectable and their gapless properties are protected by nontrivial topology. This scenario is consistent with our experimental observations as well as previous STM results on Bi crystal [23,24] but eliminates the contradictions of topological phase transition above 8 BL [17].

### D. Double-BL Bi(111) edge

Having identified the 1D TESs on the single-BL edge, we now come to the double-BL edge. Figure 10(a) shows a double-BL step of the Z-edge with a height of 0.79 nm supported on a 7 BL Bi film, the structural sketch of which is illustrated in Fig. 10(b). On this edge, a localized TES is also clearly observed (Fig. 10(c)). By comparing the STS spectra taken on a 7 BL Bi terrace and at the Z-edge (Figs. 10(d) and 10(e)), we find that the spectroscopic difference (marked by arrows) on and far away from the edges looks very similar for single- and double-BL edge. This is consistent with previous DFT calculations that predict the double-BL Bi(111) thin film to be topologically nontrivial [16]. Thus, we believe that the detected STS peak originates from the contribution of the TES. To our knowledge, this observation gives the first experimental evidence that conforms to the theoretically predicted nontrivial topology of double-BL Bi(111) thin film, without the even-odd oscillation of topological triviality as early perceived [8]. It highlights the significant role of strong inter-BL coupling in defining the unique topological properties of Bi(111) thin films.

### IV. CONCLUSION

In conclusion, we have investigated the topological properties of single-BL and double-BL Bi supported on multi-BL Bi(111) film substrate. STS measurements on the islands recorded at different thickness of the underlying Bi film (from 4 to 9 BL) show very similar fingerprints of 1D edge states. By combining these results with DFT calculations and a rather general perturbation analysis, we conclude that there are TESs of the single-BL Bi independent of the thickness of the Bi substrate. The double-BL edge also hosts TES that conform to previous theoretical predictions. Besides, we find that the exact dispersion and intensity of the edge states are rather different on two types of edge structures (Z- and R-edges). The observed different edge states exemplify the well-known fact that the topological analysis does not predict any details of the edge state dispersion or, in turn, it shows the robustness of the presence of edge states independent of the chemical or geometrical details of the edges. In addition, by probing the proximity-induced SC gap on the Bi islands, which originates from the underlying $NbSe_2$ substrate, we could demonstrate that both types of edges are non-magnetic, since they do not induce any in-gap states. This missing magnetism is corroborated by DFT calculations, which include the substrate, and is a prerequisite for the presence ofTESs. Hence, it supports the interpretation above of the presence of TES. Our results not only provide a clue for studying the electronic topology of heterostructures composed of topological materials, but also demonstrate a viable way of manipulating the topological edge/surface states with boundary modifications that can be generalized to other TIs.

## ACKNOWLEDGMENTS

This work is funded by the National Key Research and Development Program of China (Grants No. 2017YFA0403501, No. 2016YFA0401003, and No. 2018YFA0307000), the National Science Foundation of China (Grants No. 11774105, No. 11504056, No. 11522431, and No. 11474112), and the Fundamental Research Funds for the Central Universities (Grant No. 2017KFXKJC009). P. Z. T and S.-C. Z acknowledge the Department of Energy, Office of Basic Energy Sciences,

Division of Materials Sciences and Engineering, under Contract No. DE-AC02-76SF00515. A. R. acknowledges financial support from the European Research Council (Grant No. ERC-2015-AdG-694097) and European Union's H2020 program under Grant No. 676580 (NOMAD). The project leading to this application has received funding from the European Union's Horizon 2020 research and innovation programme under Marie Sklodowska-Curie Grant No. 793609.

**Figure captions:**

Figure 1 (a) Topographic image of various thicknesses of Bi films containing both Bi(110) and Bi(111) surfaces as labeled (150 × 150 nm², $V_{bias}$ = +3.0 V, $I_t$ = 10 pA). Top and bottom right inset are the atomically resolved image of Bi(110) on 2 BL and Bi(111) on 4 BL, respectively (5 × 5 nm², $V_{bias}$ = +100 mV, $I_t$ = 100 pA). (b) The corresponding line profile across Bi edges by the cyan line in (a). The height of a single-BL step is 0.39 nm. (c) STS spectra recorded on different locations in (a). Spectra are shifted vertically for a clearer view. Data were acquired at 78 K.

Figure 2. (a) Schematics of Bi(111) lattice with two nonequivalent sublattices by magenta and cyan, respectively. (b) The differential STM morphology of Bi(111) thin films with a single-BL pit on NbSe₂ substrate (40 × 40 nm², $V_{bias}$ = +3.0 V, $I_t$ = 10 pA). Inset is the atomically resolved image of Bi(111) (5 × 5 nm², $V_{bias}$ = +100 mV, $I_t$ = 100 pA). (c),(d) Topographic image of Z-edge (R-edge) marked by black (red) square in (b), respectively (4 × 4 nm², $V_{bias}$ = +100 mV, $I_t$ = 100 pA). Inset in (d) is a zoom-in image for R-edge (1 × 2 nm², $V_{bias}$ = +100 mV, $I_t$ = 100 pA). (e) Line profiles along Z- (black) and R-edge (red) labeled in (c) and (d). (f) The calculated band structure for freestanding single-BL nanoribbon with 2×1 edge reconstruction. The shadowed regions are for the bulk states and the red lines around $E_F$ are for the edge states. The $E_F$ is set to be zero. (g),(h) The atomic model of Z- and R-edge on single-BL Bi(111) boundaries supported on substrates, respectively. (i),(j) The corresponding simulated STM images of Z- and R-edge, respectively. The unit of the isovalue is electrons per cubic angstrom.

Figure 3. The geometric structures for freestanding single-BL Bi(111) nanoribbons with (a)

reconstructed edges (R-edge), (b) perfect zigzag edge (Z-edge), and (c) relaxed Z-edge. (d) The band structure for the freestanding nanoribbons with relaxed Z-edge with SOC effect. (e),(f) The band structures for the spin-up and spin-down states of the freestanding nanoribbons with relaxed Z-edge without SOC, respectively. The in-gapped states are marked by $E_1$ and $E_2$. (g),(h) The charge densities of $E_1$ and $E_2$ states at the Γ point for spin-up and spin-down, respectively. The isovalue is 0.002 e/Å$^3$. The green box marks the calculated results for the nanoribbon with relaxed Z-edge. The shadowed regions are for the bulk states and dotted lines are for the edge states around $E_F$. The $E_F$ is set as zero.

Figure 4. (a) Large-scale STM morphology of Bi(111) films grown on NbSe$_2$ substrate (40 × 40 nm$^2$, $V_{bias}$ = +3.0 V, $I_t$ = 10 pA). (b),(c) 2D plot of the superconducting (SC) gap perpendicular and parallel to the Z-edge, respectively. The gap on 7 BL Bi(111) is a little larger than on 8 BL due to the proximity effect. (d),(e) 2D plot of the SC gap perpendicular and parallel to the R-edge, respectively. (f) Selected SC gaps measured on 7 and 8 BL Bi, as well as at the Z- and R-edge, respectively. The measured temperature is 0.4 K, and the amplitude of modulation voltage is 50 μV.

Figure 5. (a) STM image of single-BL Bi(111) step (40 × 15 nm$^2$, $V_{bias}$ = +1.0 V, $I_t$ = 10 pA). (b) d$I$/d$V$ spectra on the 7 and 8 BL surface, as well as on the upper and lower edge of the step marked in (a). arrows are elucidated in the main text. Spectra are shifted vertically. (c),(d) 2D plot of tunneling spectra across the Z- and R-edge, respectively. The cyan dashed lines mark the upper edge of the Bi step. (e),(f) 2D plot of tunneling spectra along the upper edge of the step for the Z-

and R-edge, respectively.

Figure 6. (a) Profile extracted from Fig. 5(c) along the black dotted line at 200 mV, showing the distribution of the edge state in real space for the Z-edge, with a spatial extension of 3.5 nm. (b) Profile extracted from Fig. 5(d) along the red dotted line at 90 mV, showing the distribution of the edge state in real space for the R-edge, with a spatial extension of 1.5 nm. (c) Energy dispersion relation of the edge states along the Z-edge by 1D Fourier transformation. Two scattering interference wavevectors ($q_1$ and $q_2$) with high intensities are marked by the yellow and red dashed lines, respectively, as guides to the eye. The $q_1$ signals are suppressed at lower energies. (d) The same as (c) but performed on the R-edge.

Figure 7. (a) Topographic image of single-BL Bi step supported on a 7 BL Bi film containing both Z- and R-edges as labeled (35 × 35 nm$^2$, $V_{bias}$ = 0.5 V, $I_t$ = 10 pA). (b)-(d) The simultaneous STS mappings acquired at different bias set point: $I_t$ = 100 pA and 100 mV (b), 120 mV (c), 180 mV (d), respectively. Inset in (d) is a zoom-in view of 10 × 10 nm$^2$. (e) STM image of Bi(111) film containing both Z- and R-edges (45 × 37 nm$^2$, $V_{bias}$ = +300 mV, $I_t$ = 100 pA). (f),(g) Spectroscopic mapping of the imaged area in (e) at +100 and +200 mV, respectively.

Figure 8. (a) A series of d$I$/d$V$ spectra recorded on the Bi terrace from 4 to 10 BL. Spectra are shifted vertically. (b) TESs on the Z-edge of single-BL Bi with various thicknesses of underlying Bi film substrate, which are extracted from the VHS peak positions at each edge by downward arrows. (c) The energy evolution for the single-BL TES on the Z-edge (red curve) with various

thicknesses of underlying Bi substrate (black curve). The black dashed line is the bulk Bi value of VHS (213 mV), and the red dashed line presents the TES from previous reports (183 mV) [23,43]. (d) Large-scale STM morphology of Bi(111) films including both 9 and 10 BL (30 × 30 nm$^2$, $V_{bias}$ = +3.0 V, $I_t$ = 10 pA). (e) 2D plot of tunneling spectra across both the Z- and R-edge labeled by the red line in (d). The localized TES are marked by the dashed magenta and black lines right at the Z- and R-edge, respectively.

Figure 9. (a) Schematic diagram for Bi(111) thin films including the top single-BL and underlying substrate. (b),(c) Band evolutions for the system containing the top single BL and Bi substrates with different thicknesses. The panels in the shadow of cyan and red color demonstrate the band edges for the isolated top single-BL and Bi substrate. The cyan and red lines stand for the conduction- and valence-band edge at the Γ point for the top single BL (Bi substrate), marked as "CBE@Γ" and "VBE@Γ". The middle panel without shadow shows the related energy positions at the Γ point for the entire system when interlayer coupling is fully considered. The cyan and red solid lines represent the energy level mainly contributed by the top single-BL and multi-BL substrate. The black dashed lines are $E_F$. The thickness of the Bi substrate is 3-7 BL for (b) and larger than 8 BL for (c). (d) The calculated band structures for Bi(111) thin films with different thicknesses (4-9 BL), that contain both the top single-BL and underlying Bi substrate (3-8 BL). The $E_F$ is set as zero. Insets are the zoom-in view of band structures around $E_F$ and Γ points.

Figure 10. (a) Topographic image of Bi(111) films containing both single- and double-BL boundaries (30 × 30 nm$^2$, $V_{bias}$ = +1.0 V, $I_t$ = 10 pA). The corresponding line profile of double-BL

(0.79 nm) across the edge is shown by the magenta line. (b) Structural sketch of the double-BL edge on a 7 BL Bi substrate. (c) 2D plot of tunneling spectra measured along the red dashed line in (a) crossing the double-BL edge. (d) STS spectra on a 7 BL Bi terrace (solid curve) and at the single-BL Z-edge (dashed curve), respectively. (e) STS spectra on a 7 BL Bi terrace (solid curve) and at the double-BL Z-edge (dashed curve), respectively. The arrows in (d) and (e) mark the positions of VHSs for each spectrum.

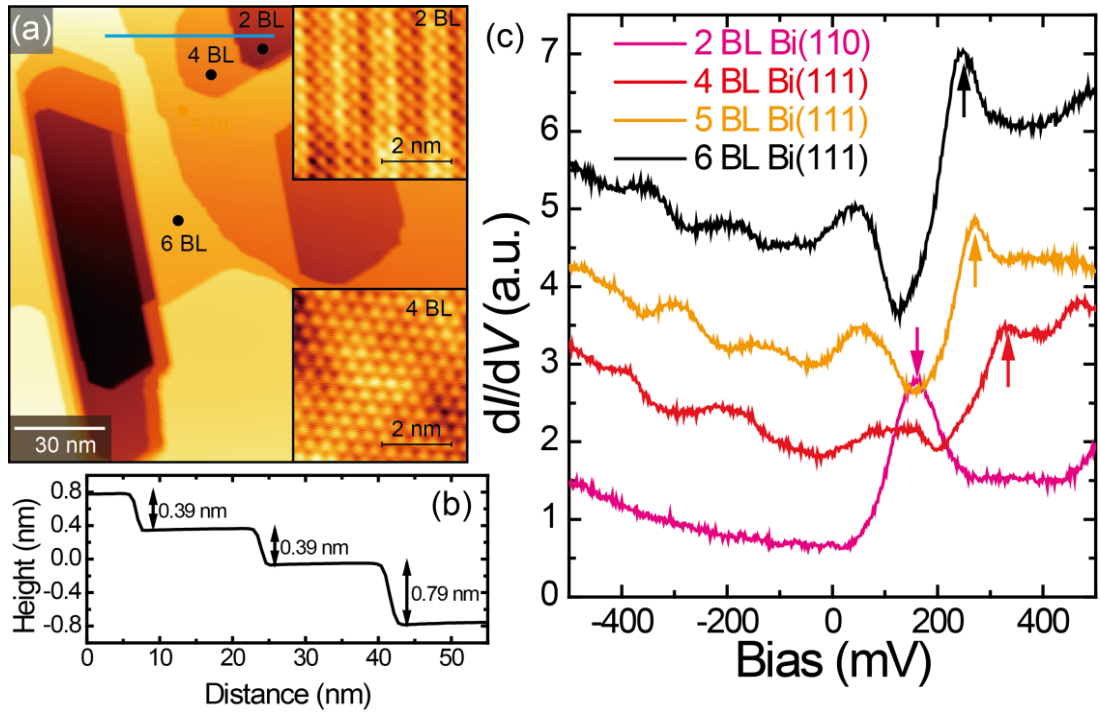

Figure 1

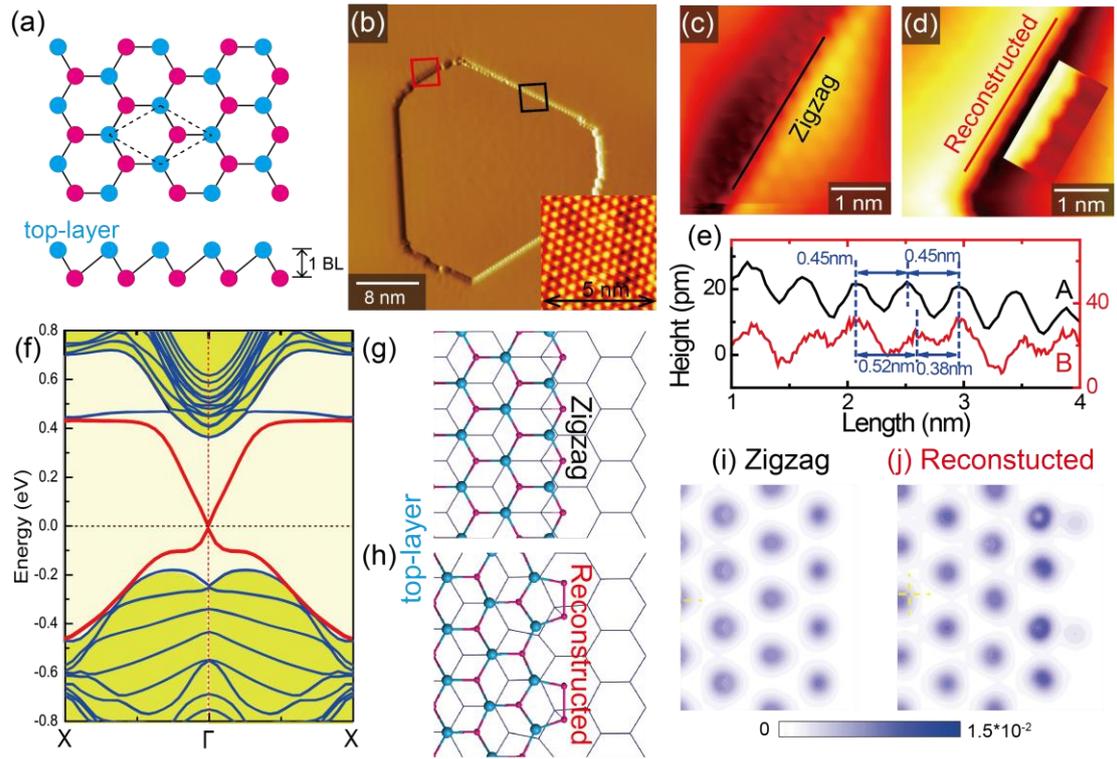

Figure 2

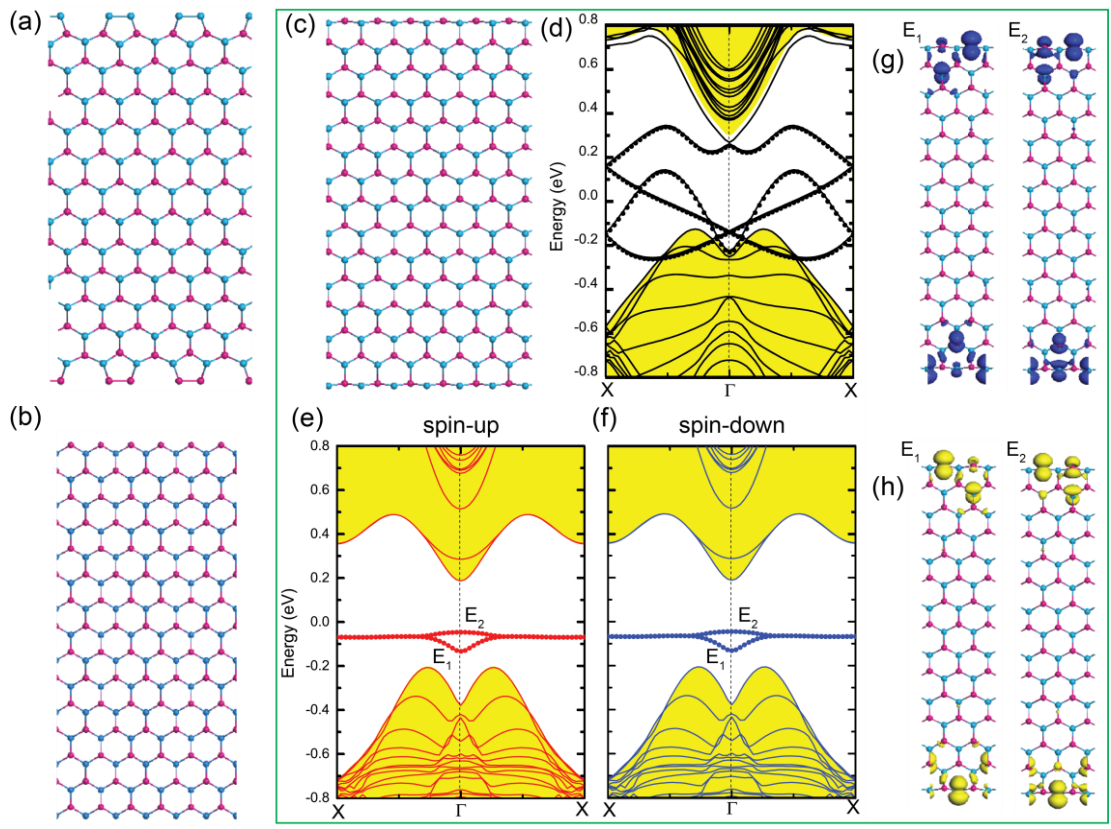

Figure 3

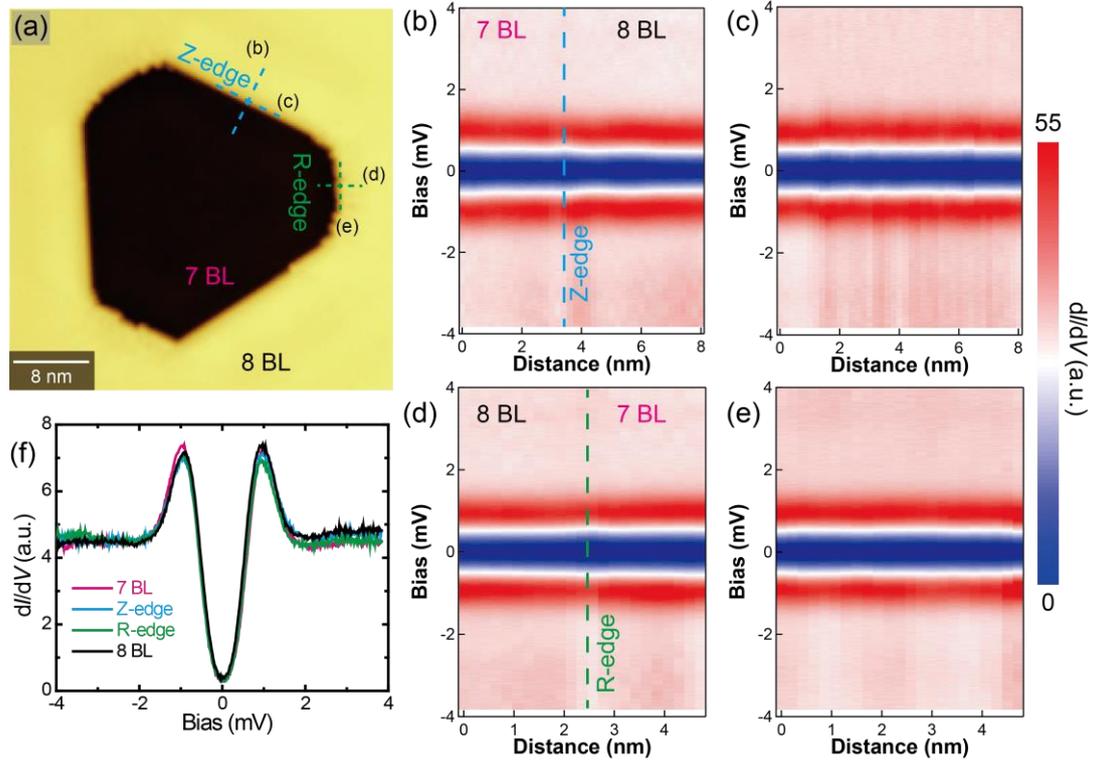

Figure 4

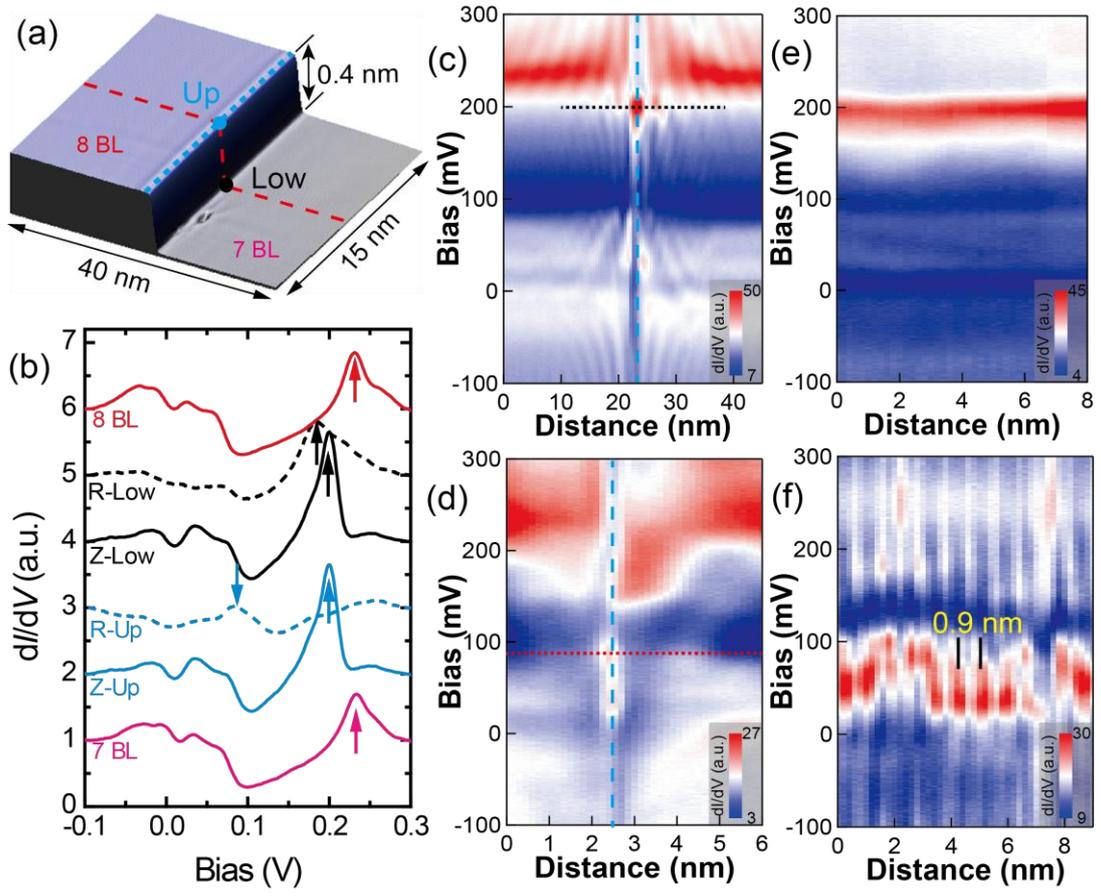

Figure 5

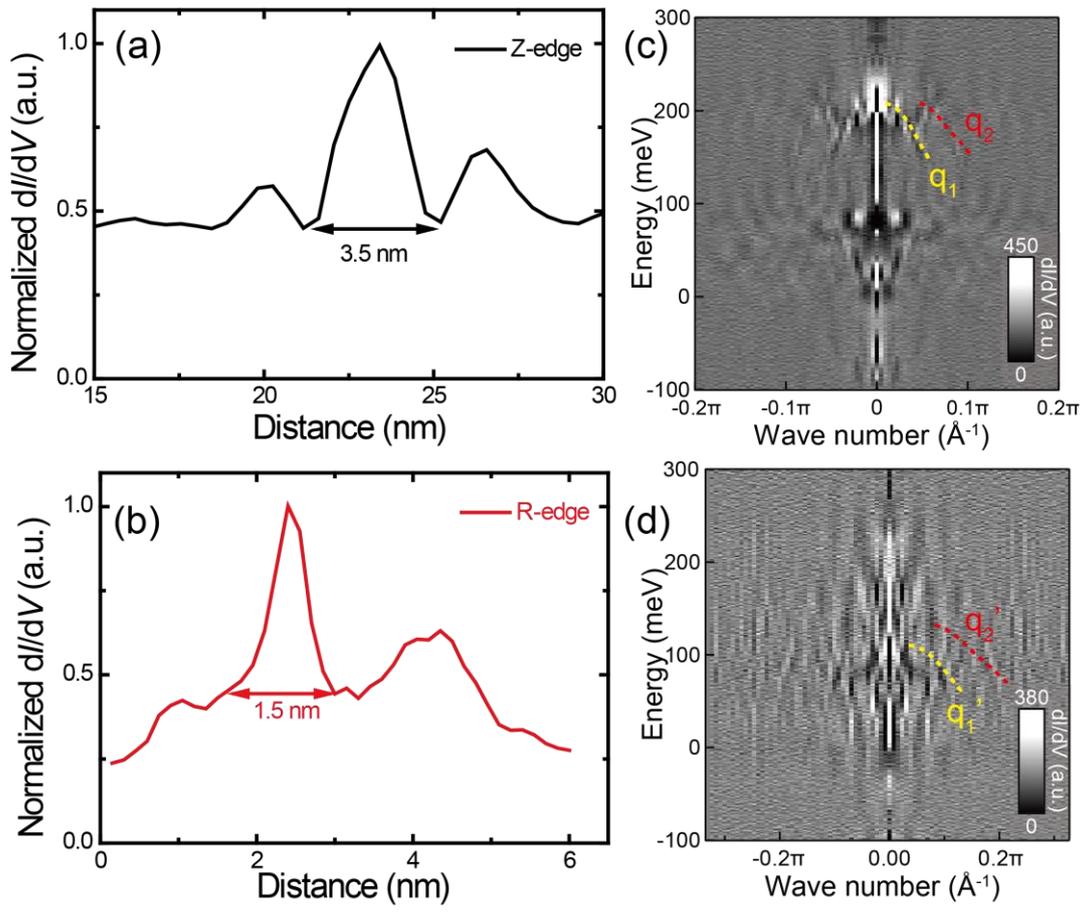

Figure 6

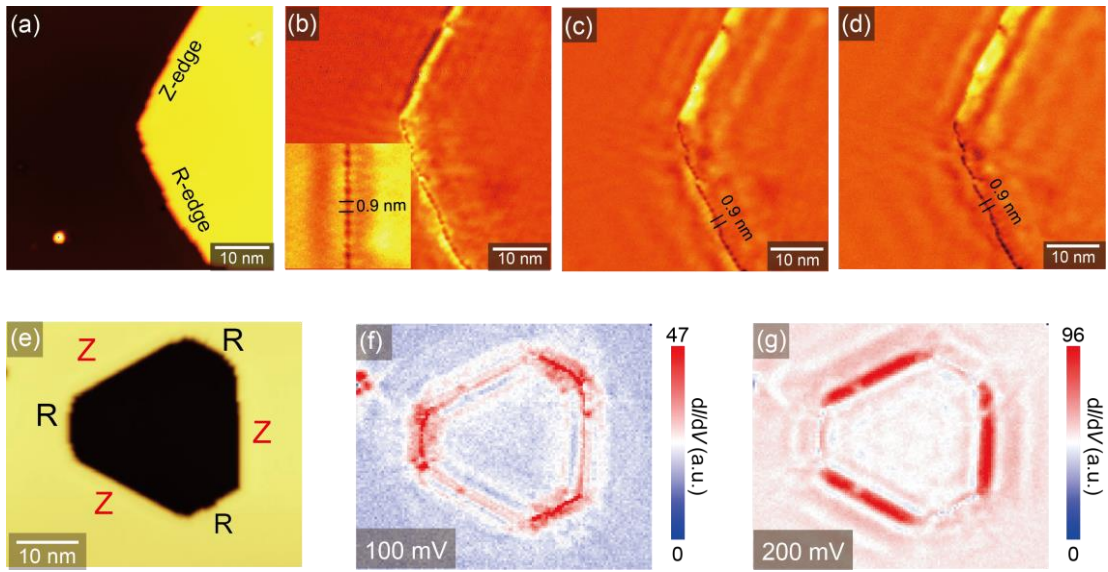

Figure 7

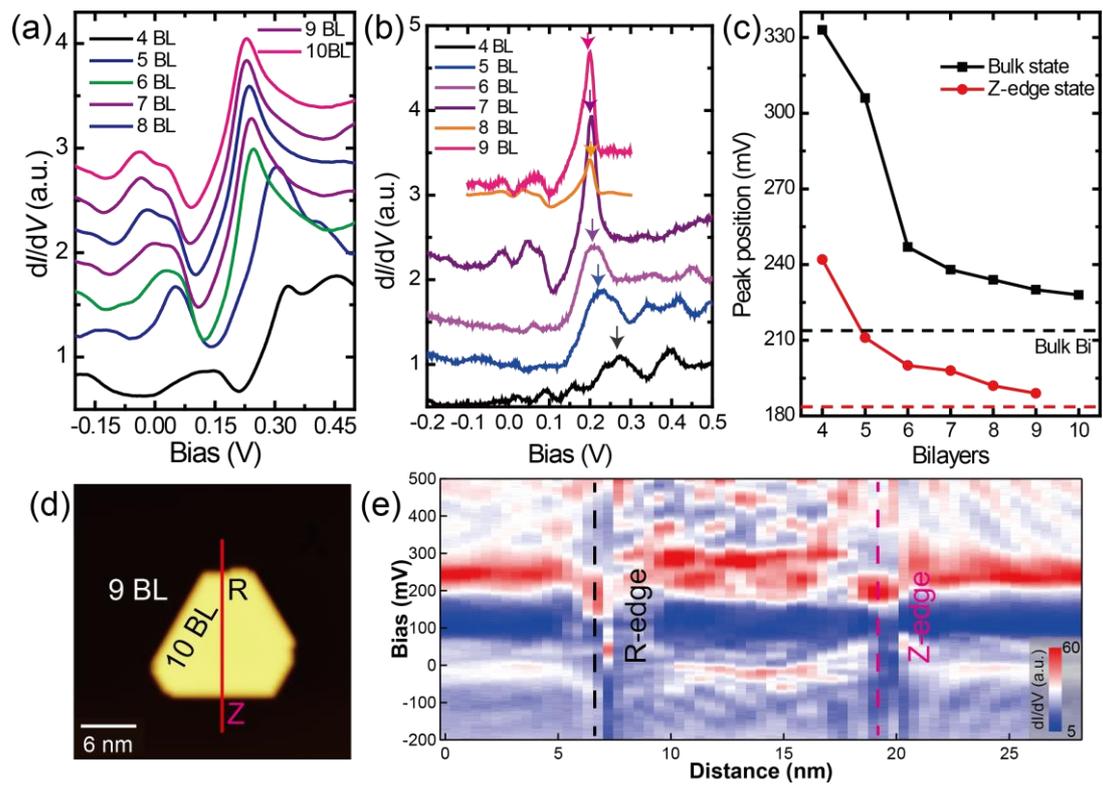

Figure 8

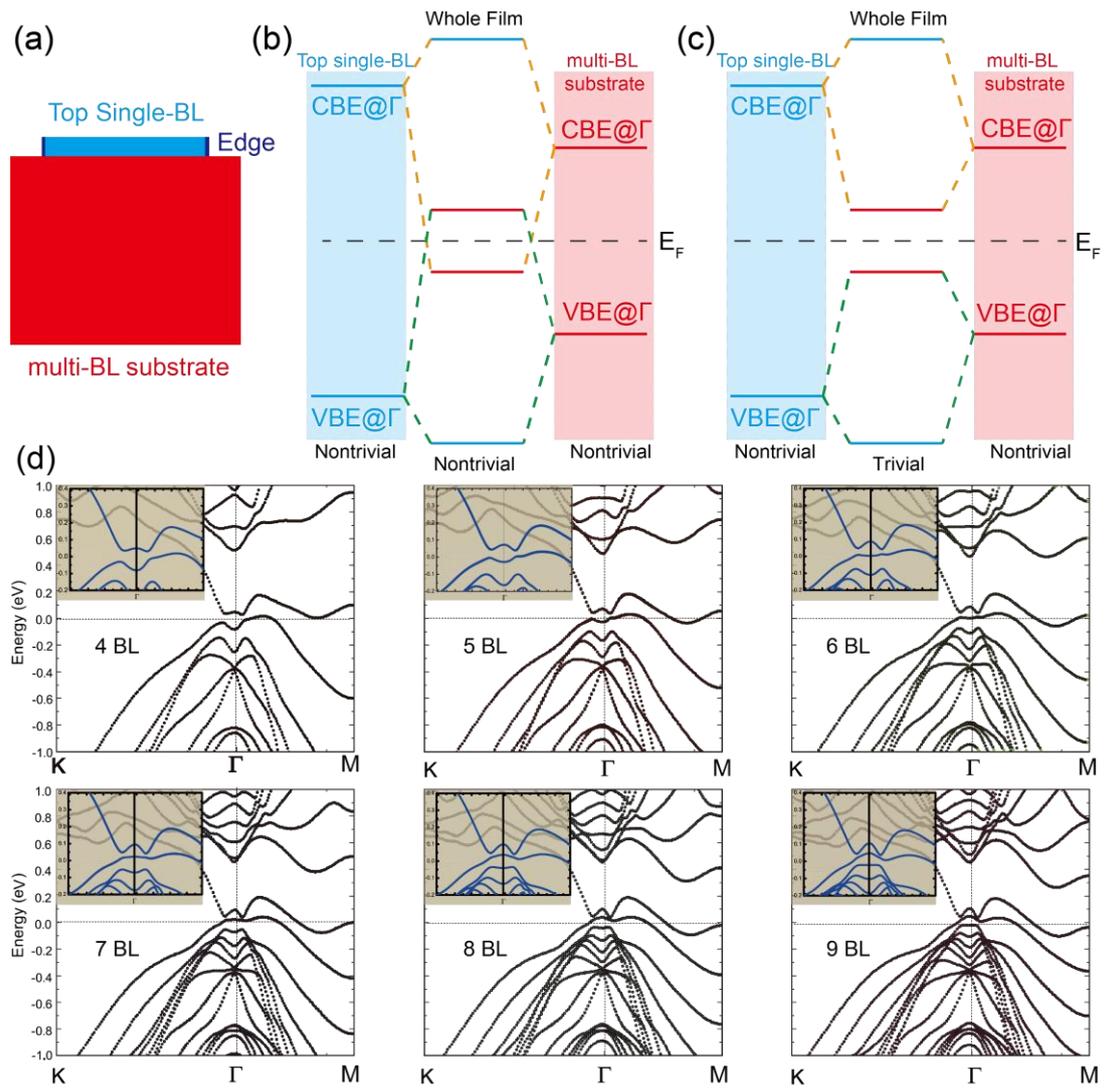

Figure 9

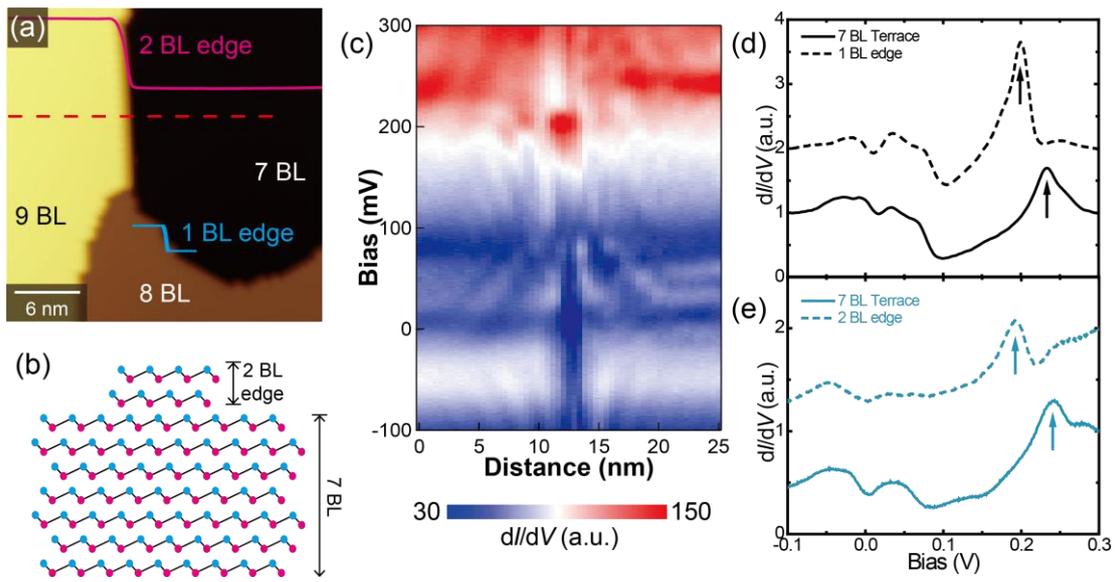

Figure 10